\documentclass[11pt]{article}

\usepackage{xr-hyper}
\usepackage[T1]{fontenc}
\usepackage{newtxtext,newtxmath}
\usepackage[margin=1in]{geometry}
\usepackage{tabularx}
\usepackage{graphicx}
\usepackage{booktabs}
\usepackage{longtable}
\usepackage{xcolor}
\usepackage{hyperref}
\usepackage{microtype}
\usepackage{float}
\usepackage[super,comma,sort&compress]{natbib}
\usepackage{tikz}
\usepackage{pgfplots}
\pgfplotsset{compat=1.18}
\usepackage{listings}
\usepackage{authblk}
\usepackage{arydshln}

\usepackage{setspace}
\usepackage{tabularx}
\newcolumntype{C}{>{\centering\arraybackslash}X}

\usepackage{caption}
\DeclareCaptionLabelSeparator{bar}{. }
\usepackage[title,titletoc]{appendix}

\newcommand{\beginsupplement}{%
  \setcounter{section}{0}%
  \renewcommand{\thesection}{S\arabic{section}}%
  \setcounter{table}{0}%
  \renewcommand{\thetable}{S\arabic{table}}%
  \setcounter{figure}{0}%
  \renewcommand{\thefigure}{S\arabic{figure}}%
  \setcounter{equation}{0}%
  \renewcommand{\theequation}{S\arabic{equation}}%
}

\hypersetup{
  colorlinks=true,
  linkcolor=blue!60!black,
  citecolor=blue!60!black,
  urlcolor=blue!60!black,
}

\title{\textbf{DNA storage approaching the information-theoretic ceiling}}

\author[1,2]{James L. Banal}

\affil[1]{Cache DNA, Inc., 1030 Brittan Avenue, San Carlos, CA 94070 USA}
\affil[2]{Correspondence should be addressed to james@cache-dna.com}

\date{}

\begin{document}
\maketitle

% ============================================================
\begin{abstract}
Synthetic DNA approaches 227.5 exabytes per gram of storage density with stability over millennial timescales. Realising this capacity requires error-correction codes that recover data from substantial synthesis and sequencing errors. Existing codecs convert noisy sequencer output into discrete base calls before error correction, discarding probabilistic information about which positions are reliable. Here we present a coding scheme that retains the sequencer's per-position posterior distributions through an integrated decoder of profile hidden Markov model alignment, log-product fusion across reads, and ordered-statistics decoding. On the DT4DDS channel simulator, the codec recovers 155.8 and 25.9 exabytes per gram of dsDNA under high- and low-fidelity conditions, exceeding the highest prior-art density on each channel by 11 and 52 percent. Under a single-encode-then-degrade protocol mapped to depurination kinetics at 25~\textdegree C in the dry state, the codec projects 282 years of decodable storage at 17.1 exabytes per gram. These results place DNA storage density within reach of the Shannon bound of the underlying channel.
\end{abstract}

% ============================================================
%\section{Introduction}
%\label{sec:intro}

DNA is an attractive medium to store digital information. It has a theoretical information density on the order of $10^9$ GB per cubic millimeter\cite{zhirnov2016}, durability measured in millennia under suitable preservation\cite{grass2015}, and enzymatic replication machinery that copies information at marginal cost. On a mass basis, the information-theoretic ceiling is 227.5 exabytes per gram of double-stranded DNA at 2 bits per base pair. Realised densities remain far below this ceiling. The most recent systematic benchmark across six representative codecs places the feasible density under realistic synthesis and sequencing conditions at 117 exabytes per gram\cite{gimpel2026}, approximately half of the information-theoretic ceiling.

To locate the gap, we examined the information lost between the sequencer's output and the decoder's input. The DNA channel carries two noise sources with different statistical structure. Synthesis errors are introduced when each oligo is written and persist in every copy of that strand thereafter. Sequencing errors are independent across reads. The sequencer emits a posterior probability distribution at each position of each read, which carries the information needed to distinguish the two. Most prior decoding pipelines discard this distribution before any error correction is attempted. Hard-decisioning a calibrated posterior strictly reduces mutual information with the true symbol, and the loss compounds across thousands of positions per strand and across thousands of strands per file.

Prior codecs take one of two architectural paths, and the dominant members of both discard soft information (Table~\ref{tab:codec_comparison}). The canonical path clusters reads of the same strand, computes a hard-symbol consensus, and feeds the consensus into the inner code. Most codecs on this path use Reed-Solomon inner codes\cite{grass2015,organick2018,mgcplus2026}. DNA Fountain uses an Luby transform fountain outer code without an inner code and relies on per-strand hash screening instead\cite{erlich2017}. Consensus averages sequencing errors across reads, which cancel in the limit, but does nothing for synthesis errors, which persist in every copy. The second path bypasses consensus by decoding each read independently. HEDGES performs a greedy tree search over insertion, deletion, and substitution hypotheses against a convolutional inner code\cite{press2020}. DNA-Aeon decodes each read with a stack-algorithm arithmetic decoder that prunes states using an internally-computed Fano metric\cite{welzel2023}. Single-read decoders avoid consensus but do not consume the per-position posterior probabilities the sequencer emits. A third line of work has propagated basecaller-derived quality information into the inner decoder\cite{chandak2020,jeon2024}, but under the assumption of pre-aligned reads and without cross-read posterior fusion, leaving information on the table at alignment and aggregation stages.

Here we present Mahoraga, a DNA storage codec that preserves soft information from sequencer to decoded byte. Raw reads bypass clustering and consensus. Each read is scored against reference strands by a profile hidden Markov model whose forward-backward pass yields per-position posterior distributions jointly with read-to-reference alignment, and posteriors across reads of the same strand are combined by log-product into a soft log-likelihood ratio for each bit. A progressive-edge-growth LDPC inner code is decoded from the soft LLRs by ordered-statistics decoding, with CRC-32 flagging inner failures as erasures for a Reed-Solomon outer code over GF($2^{16}$). The pipeline takes no hard decision until the LDPC stage, and the outer code receives both errors and erasures rather than errors alone. On a matched-parity benchmark, the codec exceeds DNA-Aeon by 1.42-fold at directly comparable cells on both the high- and low-fidelity channels and exceeds every prior codec at every tested operating point where they decode.

\begin{table}[h]
\centering
\caption{Comparison of DNA storage codec architectures. Each column describes the decoding strategy used by the codec at each stage of the pipeline.}
\label{tab:codec_comparison}
\small
\setlength{\tabcolsep}{4pt}
\renewcommand{\arraystretch}{1.2}
\begin{tabular}{l*{6}{>{\raggedright\arraybackslash}p{0.122\linewidth}}}
\toprule
Parameter & DNA-RS \cite{grass2015} & Fountain \cite{erlich2017} & HEDGES \cite{press2020} & DNA-Aeon \cite{welzel2023} & MGC+ \cite{mgcplus2026} & Mahoraga (this work) \\
\midrule
Read aggregation & Hard consensus & Hard consensus & Hard single-read & Hard single-read & Hard consensus & Soft per-read posteriors \\
\midrule
Inner code       & RS & None & Convolutional & Arithmetic & RS & LDPC with OSD \\
\midrule
Outer code       & RS & LT fountain & RS(255,223) & Raptor fountain & RS & RS GF($2^{16}$) \\
\midrule
Indel handling   & Cluster + MSA consensus & Strand dropout & Tree search & Sequential decoder & Cluster + MSA consensus & HMM posteriors \\
\bottomrule
\end{tabular}
\end{table}

\section{Results}
\label{sec:results}

\subsection{Robustness across high- and low-fidelity channels}
\label{sec:results_robustness}

\begin{figure}[H]
\centering
\includegraphics[width=\textwidth]{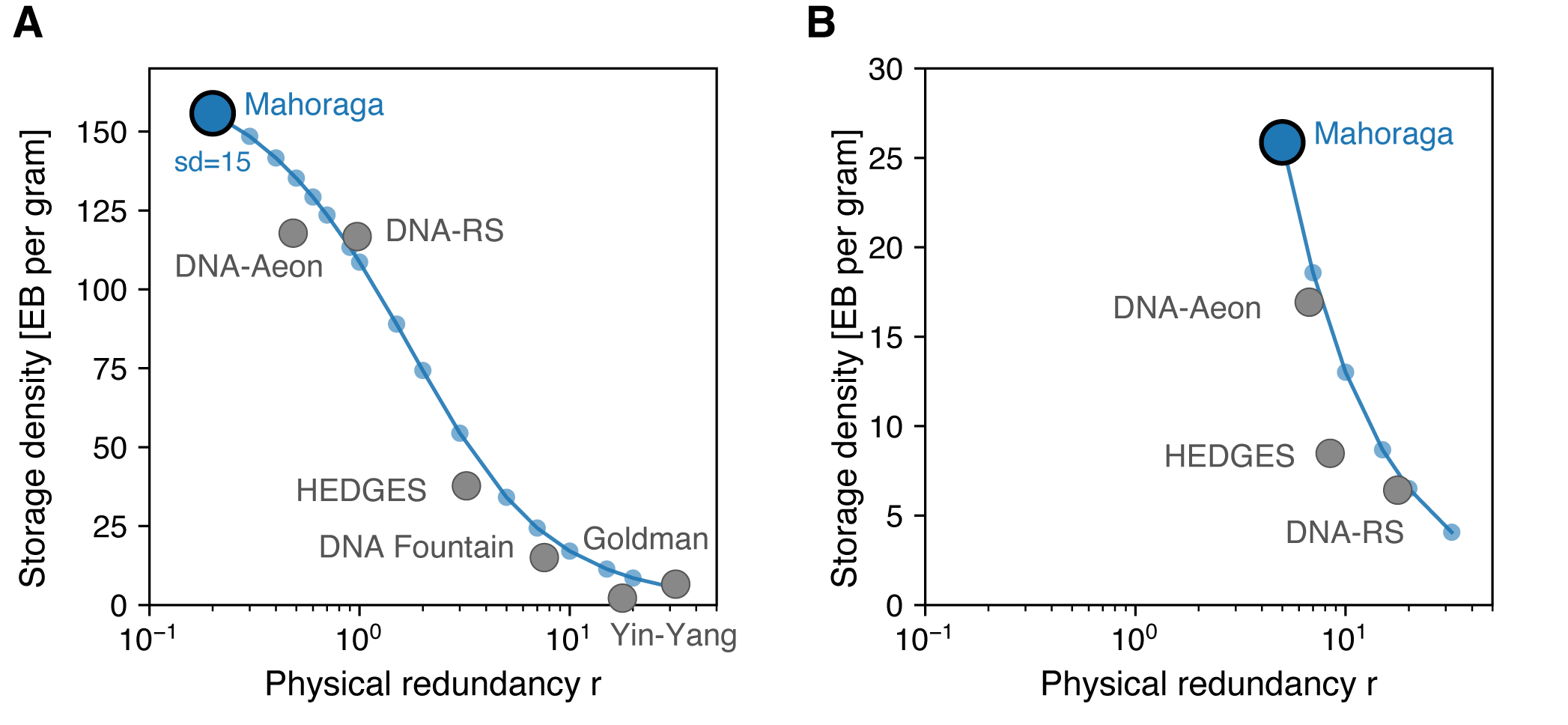}
\caption{Storage density achieved by Mahoraga compared to prior codecs on the DT4DDS channel simulator~\cite{gimpel2026}. (\textbf{A}) High-fidelity channel (Twist synthesis, Q5 PCR). (\textbf{B}) Low-fidelity channel (CustomArray synthesis, Taq PCR). Each gray marker denotes a prior codec at its peak density operating point at 30$\times$ sequencing depth, at the decoding success threshold published by the reference benchmark. The blue curve traces Mahoraga's Pareto envelope across physical redundancies at which 30 of 30 benchmark trials decode without error, with the emphasized marker indicating the peak. Mahoraga achieves 155.8 EB per gram of dsDNA on the high-fidelity channel at 15$\times$ sequencing depth and 25.9 EB per gram of dsDNA on the low-fidelity channel at 30$\times$ sequencing depth, at lower physical redundancy than any prior codec on either channel. DNA Fountain, Goldman, and Yin-Yang did not reach the decoding threshold on the low-fidelity channel.}
\label{fig:dt4dds_pareto}
\end{figure}

Mahoraga was benchmarked on the DT4DDS channel simulator~\cite{gimpel2024} using the same high-fidelity parameters (Twist synthesis, high-fidelity PCR, per-base error $\approx 10^{-3}$) as the codec comparison of~\cite{gimpel2026}. To align with the electrochemical-synthesis strand-length constraint of that study, the payload length was fixed at 126 nucleotides, matching one of the three file-size and strand-length configurations used in the associated in vitro experiments. A file of 19,456 bytes was encoded into 4,508 oligonucleotides, passed through the DT4DDS channel at physical redundancy $r = 0.2$ and sequencing depth of 15 reads per reference oligonucleotide, and recovered without error in all 30 independent trials. The achieved physical density was 155.8 EB per gram of dsDNA, 11\% higher than the best prior-art density of 140 EB per gram reported for DNA-Aeon at the same channel~\cite{gimpel2026} (DNA-Aeon's reference peak at $r = 0.49$, $\mathrm{sd} = 30$). The Mahoraga density was achieved at 2.4-fold lower physical redundancy ($r = 0.2$ vs $r = 0.49$) and 2-fold lower sequencing depth (15 vs 30 reads per oligonucleotide). The three reductions are independent axes of advantage and compound to substantially reduced synthesis and sequencing cost at matched density (Fig.~\ref{fig:dt4dds_pareto}A). In our cross-codec benchmark on the idsim high-fidelity channel described in Methods (Supplementary Table~\ref{tab:bench1_full}), DNA-Aeon at its authors' recommended settings reaches 30/30 decoding at $r \in \{2, 5, 10\}$ and peaks at 41 EB per gram at $r = 2$. This cross-codec result is on a different channel simulator than the reference 140 EB per gram figure and is not a replication of that benchmark cell.

Robustness across the DT4DDS parameter space was tested by applying the same codec configuration to the low-fidelity channel (CustomArray electrochemical synthesis, error-prone PCR, per-base error $\approx 1.5 \times 10^{-2}$)~\cite{gimpel2026}. At $r = 5$ and sequencing depth of 30 reads per reference oligonucleotide, the same file was recovered in all trials at a density of 25.9 EB per gram of dsDNA, 52\% higher than the best prior-art density of 17 EB per gram reported for DNA-Aeon on this channel~\cite{gimpel2026}, at 1.3-fold lower physical redundancy ($r = 5$ vs $r = 6.7$) and the same sequencing depth (Fig.~\ref{fig:dt4dds_pareto}B).

The 126 nt payload used above reflects the electrochemical synthesis limit of the reference study, not a codec-imposed constraint. To characterize how peak density responds to payload length, $L$ was swept on a simpler channel simulator that models per-base insertion-deletion-substitution noise and lognormal coverage variation without chemistry-specific biases. Peak density rises to 158.2 EB per gram of dsDNA at $L = 200$ nt ($r = 0.3$, 1,865 oligonucleotides) and 161.1 EB per gram at $L = 250$ nt ($r = 0.3$, 1,465 oligonucleotides), with diminishing returns beyond 200 nt as Reed-Solomon outer decoding complexity and per-strand dropout tolerance begin to erode the advantage. Payload-length selection is therefore a synthesis-platform question rather than a codec limitation, and the 126 nt DT4DDS result reported above is a lower bound on densities achievable at longer payload lengths available on other synthesis platforms.

\subsection{Inner-code isolation at matched parity}
\label{sec:results_matched_parity}

To bound the inner-code contribution, Mahoraga, DNA-Aeon, and MGC+ were benchmarked on the idsim channel with outer-code parity held at Mahoraga's auto-sized value at each physical redundancy. This is an equalized-budget comparison rather than a clean isolation. Competing codecs are held to a parity budget selected by Mahoraga's encoder, which in some cells is below what the competing codec would allocate when tuned. Where a competing codec fails at matched parity but succeeds at its native parity on the same cell (Table~\ref{app:bench1_full} and Table~\ref{app:bench2_full}), the gap between native and matched performance reflects that codec's own parity sensitivity rather than inner-code quality. The matched-parity density advantage reported below is therefore an upper bound on Mahoraga's inner-code advantage at cells where the competing codec fails under matched parity, and a direct measure at cells where the competing codec succeeds under both.

At the lowest redundancy where DNA-Aeon decodes under matched parity on the high-fidelity channel ($r = 0.02$), Mahoraga reaches 153.2 EB per gram of dsDNA against DNA-Aeon's 107.7 EB per gram, a 1.42-fold density advantage at a directly matched cell. MGC+ does not decode at $r = 0.02$ under any parity and its lowest matched-parity success is $r = 1.0$, where Mahoraga (97.8 EB per gram) exceeds MGC+ (59.9 EB per gram) by 1.63-fold. On the low-fidelity channel, DNA-Aeon's only matched-parity success within the swept range is $r = 0.5$ at 64.9 EB per gram, where Mahoraga reaches 92.3 EB per gram (1.42-fold). MGC+'s lofi peak under matched parity is 59.1 EB per gram at $r = 0.5$, against Mahoraga's 92.3 EB per gram at the same cell (1.56-fold). Cells where competing codecs fail under matched parity but Mahoraga succeeds (Table~\ref{app:bench2_full}) are reported as architectural envelope differences rather than as density ratios.

The matched-parity gap is a direct measure of the information advantage the inner code extracts per read. With outer codes equalized, the density difference reflects only the inner-code and read-aggregation architecture. DNA-Aeon's stack-algorithm arithmetic decoder operates on hard symbols with an internally-computed Fano metric, while Mahoraga's HMM posteriors carry per-position sequencer quality information into the LDPC decoder. The low-fidelity gap is larger because the per-position information advantage grows with channel error rate.

The operating point at which each codec peaks is also informative. Mahoraga and DNA-Aeon both peak at $r = 0.02$ under high-fidelity conditions, where the channel is near its dropout limit. MGC+ peaks at $r = 1.0$, one and a half orders of magnitude higher, because its CD-HIT plus majority-vote consensus requires multiple intact reads per template before inner decoding can proceed. The ability of Mahoraga to decode at $r = 0.02$ under low-fidelity conditions is the largest driver of its density advantage. At an order of magnitude less input material, effective density per gram scales inversely with redundancy.

\begin{figure}
\centering
\includegraphics[width=0.7\textwidth]{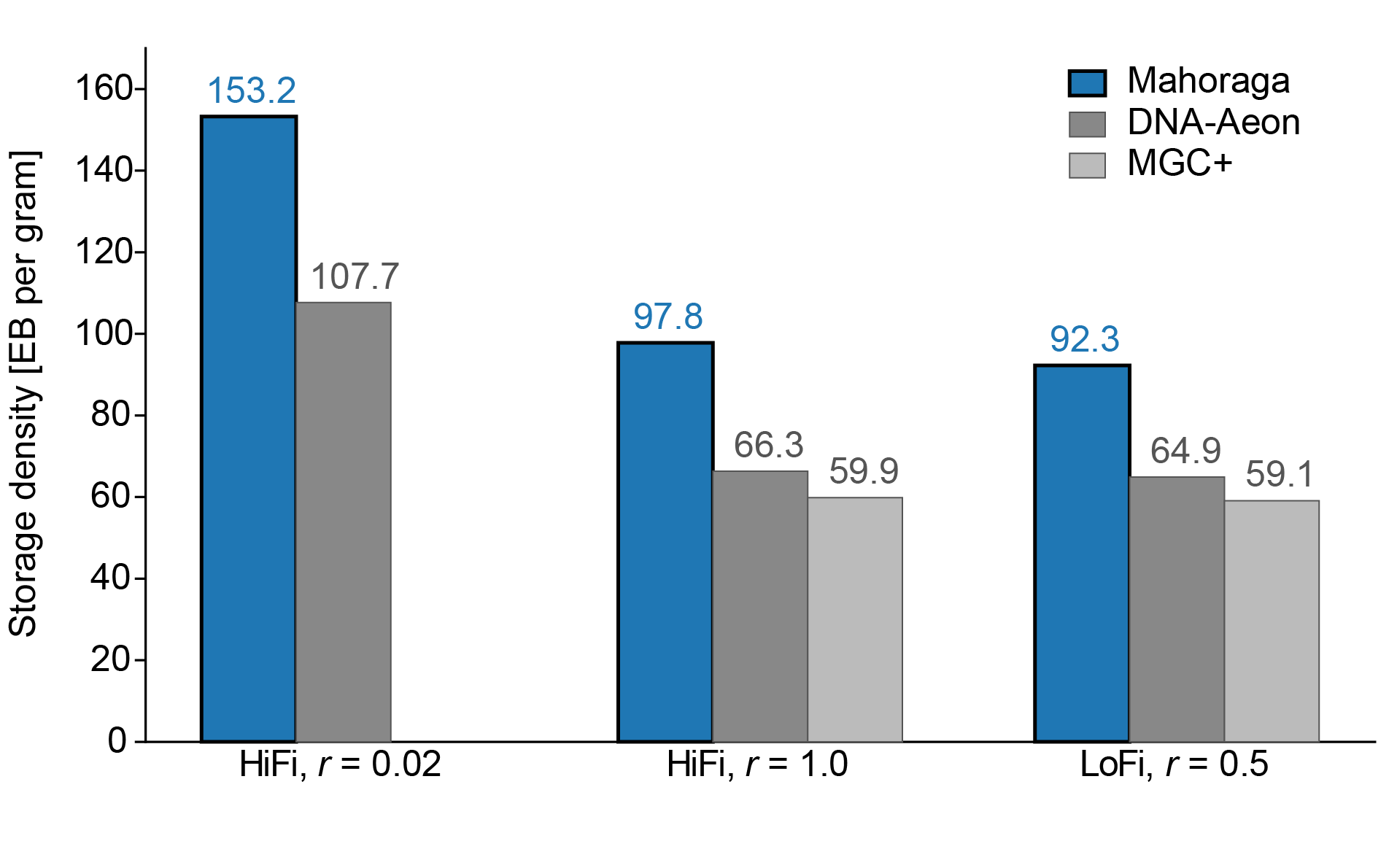}
\caption{Storage density at matched parity and matched physical redundancy. Each group plots storage density for Mahoraga and the codecs that decode at that cell under matched outer-code parity. At $r = 0.02$ on the high-fidelity channel, MGC+ does not reach 30 of 30 decoding and is omitted. Matched parity holds the outer-code parity at Mahoraga's auto-sized value per cell and is applied to all codecs. Mahoraga exceeds DNA-Aeon by 1.42-fold at $r = 0.02$ on the high-fidelity channel (153.2 against 107.7 EB per gram of dsDNA), exceeds MGC+ by 1.63-fold at $r = 1.0$ (97.8 against 59.9 EB per gram), and exceeds DNA-Aeon by 1.42-fold at $r = 0.5$ on the low-fidelity channel (92.3 against 64.9 EB per gram).}
\label{fig:matched-parity}
\end{figure}

\subsection{Performance against the alphabet ceiling}
\label{sec:shannon_limit}

Density comparisons so far have been made between codecs. An absolute reference point is the alphabet ceiling for a quaternary alphabet, which is 2 bits per base pair~\cite{shomorony2022}. The exact Shannon capacity of the noisy shuffling-sampling channel that describes DNA storage depends on dropout fraction $q_0 = e^{-r}$, per-base substitution and indel rates, and the regime parameter $\beta = L / \log_2 M$~\cite{shomorony2022}, and is in general strictly smaller than the alphabet ceiling. We use the alphabet ceiling as the benchmark because it is the only universal closed-form bound. At a given physical redundancy $r$, the maximum density any codec can achieve, accounting for stochastic dropout, is
\begin{equation}
\rho_{\mathrm{max}}(r) = 2 \cdot \frac{1 - e^{-r}}{r} \cdot \rho_{\mathrm{conv}},
\label{eq:alphabet_ceiling}
\end{equation}
where $(1 - e^{-r})$ is the Poisson survival fraction capping the outer-code rate, the factor $1/r$ accounts for the mass cost of redundant copies, and $\rho_{\mathrm{conv}} = 113.7$ EB per gram per bit per base is the dsDNA mass conversion constant (Methods). Fractions of $\rho_{\mathrm{max}}$ reported below are upper bounds on the codec's fraction of the true channel capacity, since channel-specific capacity bounds derived from concrete noise models~\cite{shomorony2022,lenz2023} are strictly smaller than the alphabet ceiling.

Mahoraga sits at 68\% of the alphabet ceiling on the high-fidelity channel and 52\% on the low-fidelity channel, constant to within 0.1 percentage point across every successful operating point in the benchmark (Table~\ref{tab:shannon_fraction}; the full cell-by-cell listing across all ten redundancies on both channels is given in Supplementary Table~\ref{app:shannon_fraction_full}). DNA-Aeon operates at 46 to 48\% of the ceiling on the high-fidelity channel. MGC+ operates at 33 to 44\% across channels and redundancies. No prior codec in the matched-parity benchmark exceeds 48\% of the ceiling at any tested operating point.

\begin{table}[H]
\centering
\caption{Codec densities as a fraction of the alphabet ceiling at matched-parity operating points. $\rho_{\mathrm{max}}$ gives the alphabet ceiling in EB per gram of dsDNA, computed from Eq.~\eqref{eq:alphabet_ceiling}. Each codec column reports that codec's density at the matched-parity cell as a percentage of $\rho_{\mathrm{max}}$. Densities are from Supplementary Table~\ref{app:bench2_full}. A dash indicates the codec did not reach 30 of 30 decoding at this cell. DNA-Aeon on the low-fidelity channel decodes only at $r = 0.5$ under matched parity (where the parity budget is 0.72) and fails at higher $r$ where the matched-parity budget drops below DNA-Aeon's decoder reliability threshold at low-fidelity error rates.}
\label{tab:shannon_fraction}
\small
\setlength{\tabcolsep}{6pt}
\renewcommand{\arraystretch}{1.15}
\begin{tabular}{llrrrr}
\toprule
Channel & $r$ & $\rho_{\mathrm{max}}$ [EB per g of dsDNA] & Mahoraga & DNA-Aeon & MGC+ \\
\midrule
High-fidelity & 0.02 & 225.1 & 68\% & 48\% & --- \\
High-fidelity & 1.0  & 143.7 & 68\% & 46\% & 42\% \\
High-fidelity & 5.0  &  45.2 & 68\% & 48\% & 43\% \\
High-fidelity & 10.0 &  22.7 & 68\% & 48\% & 44\% \\
\midrule
Low-fidelity  & 0.5  & 178.9 & 52\% & 36\% & 33\% \\
Low-fidelity  & 1.0  & 143.7 & 52\% & --- & 39\% \\
Low-fidelity  & 5.0  &  45.2 & 52\% & --- & 42\% \\
\bottomrule
\end{tabular}
\end{table}

Mahoraga's alphabet-ceiling fraction is 68\% on the high-fidelity channel and 52\% on the low-fidelity channel at every successful operating point in the benchmark (to within 0.1 percentage point). This near-exact constancy reflects that the fraction factors into the codec's structural rate and its decoder's information-extraction efficiency, both of which are properties of the architecture rather than the operating point. Writing the fraction as
\begin{equation}
\frac{\rho_{\mathrm{Mahoraga}}}{\rho_{\mathrm{max}}(r)} = \eta_{\mathrm{code}} \cdot \eta_{\mathrm{FBL}} \cdot \eta_{\mathrm{ch}},
\label{eq:decomposition}
\end{equation}
the three factors are the codec structural rate $\eta_{\mathrm{code}} = \eta_{\mathrm{inner}} \cdot s^{-1}$, the inner-decoder finite-blocklength rate as a fraction of infinite-blocklength binary-channel capacity at block length 252 ($\eta_{\mathrm{FBL}}$), and a residual that absorbs the gap between alphabet ceiling and the channel-specific per-base capacity at the operating noise level ($\eta_{\mathrm{ch}}$). The 252-bit inner codeword length is common to both channels, but the LDPC check-node degree differs by design. The high-fidelity channel uses a $(d_v=3,\, d_c=84)$ PEG construction, yielding an inner information budget of $k = 243$ bits. The low-fidelity channel uses $(d_v=3,\, d_c=21)$, yielding $k = 216$ bits. After a 32-bit CRC and alignment to even bytes, the per-oligo user payload is 208 bits on the high-fidelity channel and 176 bits on the low-fidelity channel, giving inner code rates $\eta_{\mathrm{inner}} = 208/252 \approx 0.825$ on high-fidelity and $\eta_{\mathrm{inner}} = 176/252 \approx 0.698$ on low-fidelity. The PEG parity-check matrices carry more rows than their rank (9 rows of rank 3 on high-fidelity, 36 rows of rank 12 on low-fidelity), with the additional rows contributing redundant parity-check equations that enrich the Tanner graph for belief propagation without consuming codeword bits. The outer safety margin $s^{-1} = 0.926$ is channel-independent (Mahoraga sizes outer parity at $1.08\times$ the minimum needed to cover expected dropout and inner-decode failure), so $\eta_{\mathrm{code}} = 0.764$ on high-fidelity and $\eta_{\mathrm{code}} = 0.647$ on low-fidelity.

On the high-fidelity channel the observed joint residual $\eta_{\mathrm{FBL}} \cdot \eta_{\mathrm{ch}} = 0.680 / 0.764 \approx 0.89$. The Polyanskiy-Poor-Verd\'u normal-approximation bound for the binary symmetric channel at $p_{\mathrm{sub}} = 1.3 \times 10^{-3}$, $n = 252$, and target error $\varepsilon = 10^{-6}$, with the $\tfrac{1}{2}\log_2 n / n$ third-order correction~\cite{polyanskiy2010}, gives an FBL rate $R(n,\varepsilon) \approx 0.898$ bit per channel use against an infinite-blocklength binary capacity $C(p_{\mathrm{sub}}) \approx 0.986$, so $\eta_{\mathrm{FBL}} \approx 0.91$. The residual $\eta_{\mathrm{ch}} \approx 0.89 / 0.91 \approx 0.98$ on the high-fidelity channel reflects that the per-base channel capacity at the operating substitution and indel rates is within 2\% of the alphabet ceiling. A tighter ceiling would shrink this residual toward unity.

On the low-fidelity channel the joint residual is $0.515 / 0.647 \approx 0.80$. Two factors grow with channel error rate. The finite-blocklength penalty increases because the channel dispersion $V$ grows with $p_{\mathrm{sub}}$~\cite{polyanskiy2010}. Numerically, at $p_{\mathrm{sub}} = 8 \times 10^{-3}$ (the combined synthesis plus sequencing substitution rate on the low-fidelity channel) and the same $(n, \varepsilon)$, the PPV normal approximation with the third-order correction gives $R(n,\varepsilon) \approx 0.763$ against $C(p_{\mathrm{sub}}) \approx 0.933$, yielding $\eta_{\mathrm{FBL}} \approx 0.82$. The residual $\eta_{\mathrm{ch}} \approx 0.80 / 0.82 \approx 0.97$ on the low-fidelity channel shows that the decoder operates within roughly 3\% of the composite finite-blocklength-and-capacity envelope, similar to the high-fidelity margin. The 16-percentage-point difference between the 68\% high-fidelity and 52\% low-fidelity alphabet-ceiling fractions decomposes as a 12-percentage-point drop in $\eta_{\mathrm{code}}$ (from 0.76 to 0.65, reflecting the lower-rate LDPC construction Mahoraga deploys at the higher error rate) and a 9-percentage-point drop in $\eta_{\mathrm{FBL}}$ (from 0.91 to 0.82, reflecting the larger finite-blocklength penalty at higher substitution rates), partly offset by a slight increase in $\eta_{\mathrm{ch}}$ (from 0.98 to 0.97 is within rounding).

The 22-percentage-point gap between Mahoraga and DNA-Aeon at matched parity on the high-fidelity channel at $r = 1.0$ (68\% against 46\%) reflects three architectural differences specific to Mahoraga's inner decoder. A profile hidden Markov model computes per-position posteriors jointly with read-to-reference alignment, handling indels within the posterior computation rather than requiring pre-aligned reads. Posteriors from multiple reads of the same reference are combined by log-product, extracting information from coverage that per-read decoders discard. Ordered Statistics Decoding operates directly on the per-bit log-likelihood ratios derived from these posteriors, enumerating candidate codewords by flipping low-magnitude-LLR positions up to a specified order and verifying each against the CRC. DNA-Aeon filters reads on quality before the arithmetic decoder, committing to a single base call per position before outer decoding~\cite{welzel2023}. Prior soft-information codecs for DNA storage have propagated per-base quality scores into LDPC belief-propagation decoders under the assumption of pre-aligned reads~\cite{chandak2020,jeon2024}. These codecs are not in the matched-parity benchmark and would be expected to fall between DNA-Aeon and Mahoraga in alphabet-ceiling fraction, with position depending on the channel's indel rate relative to the pre-alignment assumption.

The largest single improvement available to Mahoraga would come from reducing the outer-code safety margin, which is a reliability-density trade-off under the operator's control rather than a coding-theoretic limit. The inner-code rate is bounded by the 126-nt strand length imposed by electrochemical synthesis. On synthesis platforms supporting longer strands, the alphabet-ceiling fraction rises further (Supplementary Table~\ref{app:length-scaling}).

\subsection{Storage longevity}
\label{sec:results_longevity}

The DT4DDS and matched-parity benchmarks measure decoding performance at a single instant in time. Archival storage additionally requires a codec to tolerate accumulated degradation between write and read. To quantify longevity, a single-encoding-then-degrade benchmark was constructed. A file is encoded once at initial physical redundancy $r_{\mathrm{initial}}$ with the codec's native parity allocation, and the channel physical redundancy is then swept downward to simulate progressive oligonucleotide loss. The lowest channel redundancy at which at least 29 of 30 trials still decode without error defines the decoding cliff for that encoding. Cliff values are mapped to storage time by inverting depurination kinetics at 25~\textdegree C in the dry state~\cite{lindahl1972, allentoft2012}.

At $r_{\mathrm{initial}} = 5$, Mahoraga tolerated degradation to $r = 3.25$ before the decoding cliff, corresponding to 133 years of storage at a physical density of 34.1 EB per gram of dsDNA (Table~\ref{tab:longevity}). MGC+ at the same initial redundancy tolerated degradation to $r = 3.5$ (110 years, 19.5 EB per gram). At $r_{\mathrm{initial}} = 10$, both codecs tolerated degradation to $r = 4$, at physical densities of 17.1 EB per gram for Mahoraga and 9.9 EB per gram for MGC+. DNA-Aeon failed to reach the 29-of-30 criterion at $r_{\mathrm{initial}} = 2$ even before any degradation was applied, establishing a higher floor for archival storage with that codec. At $r_{\mathrm{initial}} = 10$, the only cell in Table~\ref{tab:longevity} where DNA-Aeon reached the 29-of-30 criterion at its decoding cliff, Mahoraga delivered 1.55-fold higher density (17.1 against 11.0 EB per gram of dsDNA) at a longer projected longevity (282 against 214 years).

\begin{table}[H]
\centering
\renewcommand{\arraystretch}{1.2}
\caption{Storage density in EB per gram of dsDNA and projected longevity at 25~\textdegree C dry storage. Longevity is projected from the decoding cliff (lowest channel redundancy at which at least 29 of 30 trials decode without error) using the chemistry model described in Methods.}
\label{tab:longevity}
\begin{tabularx}{\textwidth}{lCCCC}
\toprule
 & \multicolumn{2}{c}{$r_{\text{initial}}=5$} & \multicolumn{2}{c}{$r_{\text{initial}}=10$} \\
\cmidrule(lr){2-3} \cmidrule(lr){4-5}
Codec & Density [EB per gram of dsDNA] & Longevity [y] & Density [EB per gram of dsDNA] & Longevity [y] \\
\midrule
Mahoraga & 34.1 & 133 & 17.1 & 282 \\
MGC+ & 19.5 & 110 & 9.9 & 282 \\
DNA-Aeon & -- & -- & 11.0 & 214 \\
\bottomrule
\end{tabularx}
\end{table}

\section{Discussion}
\label{sec:discussion}

The principal finding is that a DNA storage codec operating on the per-position posterior distribution emitted by the sequencer achieves higher density under matched channel conditions than a codec operating on hard-symbol sequences. On the DT4DDS channel, Mahoraga reached 155.8 EB per gram of dsDNA under high-fidelity conditions and 25.9 EB per gram under low-fidelity conditions, exceeding the peak densities reported by~\cite{gimpel2026} for each of six prior codecs on the same channel. The margin over DNA-Aeon, the highest-performing prior codec in both regimes, was 11 and 52 percent respectively. At matched outer parity at directly comparable cells, Mahoraga exceeded DNA-Aeon by 1.42-fold at $r = 0.02$ on the high-fidelity channel (153.2 against 107.7 EB per gram) and again by 1.42-fold at $r = 0.5$ on the low-fidelity channel (92.3 against 64.9 EB per gram), the only low-fidelity redundancy at which DNA-Aeon decoded under matched parity. Framed against the alphabet ceiling of 2 bits per base pair adjusted for Poisson dropout, Mahoraga operates at 68 percent of the ceiling on the high-fidelity channel and 52 percent on the low-fidelity channel, approximately twenty-two percentage points above DNA-Aeon on the high-fidelity channel and approximately nine percentage points above any prior codec on the low-fidelity channel in the matched-parity benchmark.

The information-theoretic basis for these results parallels what genomics variant callers have done for two decades. GATK HaplotypeCaller scores reads against locally-assembled candidate haplotypes via a pair hidden Markov model that consumes per-base quality scores~\cite{vanderauwera2020}, and DeepVariant encodes aligned reads into a multi-channel pileup tensor preserving per-read base, quality, mapping quality, and strand information, which is then classified into genotype likelihoods by a convolutional neural network~\cite{poplin2018}. Neither tool collapses reads to a consensus before inference. DNA storage codecs have done the opposite. Prior codecs hard-call or consensus-reduce before error correction, and the per-position posterior probabilities the sequencer emits are discarded. Hard-decisioning a calibrated posterior strictly reduces mutual information with the true symbol, and the loss compounds across thousands of positions per strand and across thousands of strands per file. Mahoraga preserves these posteriors through the profile hidden Markov model, combines them across reads of the same reference by log-product, and passes the resulting soft log-likelihood ratios to Ordered Statistics Decoding, which enumerates candidate codewords by flipping low-magnitude-LLR positions and verifying each against the CRC. A Berlekamp-Massey step in the outer decoder identifies CRC false positives that would otherwise propagate as silent errors, promoting them to erasures for Reed-Solomon interpolation. No hard decision is taken until the OSD stage, and the outer code accordingly receives both errors and erasures rather than errors alone. The density advantage observed in the benchmarks is the information-theoretic consequence of this preservation, and the gap widens on the low-fidelity channel because the information content of a posterior grows relative to that of a hard symbol as the per-position error rate rises.

A complementary consequence concerns dropout. Where prior codecs provide against dropout through molecular redundancy, Mahoraga discharges the cost through outer-code parity scaled to the survival fraction of the channel, a provision made feasible only by an inner pipeline that decodes reliably from the sparse reads that low physical redundancy affords. The regime most favourable to the design is therefore low $r$ with extreme outer parity, and it is precisely the regime in which competing codecs fail. DNA-Aeon reaches $r = 0.02$ on the high-fidelity channel but cannot maintain reliability on the low-fidelity channel below $r = 0.5$, where the per-read error rate exceeds what its stack-algorithm arithmetic decoder can recover from without consensus averaging across multiple copies. MGC+ is excluded from $r = 0.02$ altogether because its CD-HIT clustering and Kalign multiple alignment both presuppose several reads per reference, a condition that low physical redundancy does not provide.

The initial physical redundancy $r_{\mathrm{initial}}$ at which a file is encoded is an operator-selected parameter, and the trade-off between instantaneous density and degradation tolerance is therefore tuneable to the use case. A low $r_{\mathrm{initial}}$ maximises density at the expense of longevity, which is appropriate to files that are frequently read and occasionally refreshed. A high $r_{\mathrm{initial}}$ provides longevity at the expense of density, which is appropriate to write-once archival storage. Mahoraga's advantage in density at matched longevity holds across the full range of $r_{\mathrm{initial}}$ tested, and the operator accordingly selects the tier appropriate to the application without forfeiting the efficiency gain demonstrated in the matched-parity benchmark. No prior DNA storage codec exposes this design space with comparable efficiency, because the alphabet-ceiling-fraction advantage established above persists across operating points rather than depending on a favourable choice among them.

The density gain comes at a corresponding cost in information-per-base efficiency. Mahoraga encodes at 0.27 bit per nucleotide against 0.50 bit per nucleotide for DNA-Aeon, a consequence of the architectural choice to pay for dropout through outer-code parity rather than through per-strand code rate. The density advantage is accordingly realised through reduced molecular redundancy at the synthesis stage rather than through denser per-base encoding. Where synthesis cost is dominated by base count, as on array-based synthesis platforms, the lower code rate partially offsets the lower physical-redundancy requirement. Where synthesis cost is dominated by pool-level fixed costs, as in industrial-scale archival synthesis, the reduction in oligonucleotide count at lower redundancy is the governing economic factor, and the trade-off resolves in Mahoraga's favour. Archival applications at scale belong in this second regime.

Two limitations of the present work warrant acknowledgement. First, the benchmarks reported here are in silico. The DT4DDS channel simulator is calibrated against synthesis-and-sequencing experiments on Twist and CustomArray platforms~\cite{gimpel2024}, and its predictions have been validated in vitro for other codecs on the same pipeline~\cite{gimpel2026}, so the density ratios reported above inherit the simulator's validation status. A direct in vitro validation of Mahoraga on representative synthesis chemistry is the natural next step and remains outside the scope of this work. Second, the 126-nucleotide payload length used throughout the DT4DDS experiments reflects the electrochemical synthesis limit of the reference study rather than any codec-imposed constraint. Synthesis platforms supporting longer payloads would permit Mahoraga to encode at higher density, as summarised in Table~\ref{tab:length_density}, and lower storage temperatures or encapsulation chemistries would extrapolate the longevity projections upward by orders of magnitude. Neither extension is evaluated here, and both are compatible with the codec as presented.

\section{Conclusion}
\label{sec:conclusion}
The results reported here suggest that the dominant loss in DNA data storage is informational rather than chemical. Prior codecs discard the sequencer's per-position posterior before error correction, and the densities they achieve under realistic conditions are the natural consequence of that choice. Mahoraga retains the posterior through to the decoder's output, and the density and reliability follow.

\section{Methods}
\label{sec:methods}

\subsection{Codec architecture}
Mahoraga is a concatenated codec. The inner code is a progressive-edge-growth low-density parity-check (LDPC) code~\cite{hu2005} operating at the bit level within each oligonucleotide, with decoding by Ordered Statistics Decoding~\cite{fossorier1995}. The outer code is an interleaved Reed-Solomon code over GF($2^{16}$)~\cite{reed1960} operating at the symbol level across all oligonucleotides, with interleaving such that a single oligonucleotide dropout removes exactly one symbol from each interleaved codeword. A per-sequence 32-bit cyclic redundancy check (CRC-32) couples the two codes by labelling each sequence as received (CRC passes) or erased (CRC fails) before outer decoding. This labelling makes the outer code efficient. Known erasures cost one Reed-Solomon parity symbol each, whereas unknown errors would cost two, so the CRC-erasure mechanism halves the parity overhead in the common case. A Berlekamp-Massey step in the outer decoder catches CRC false positives that slip past the inner-code check at a cost of two Reed-Solomon parity symbols per corrected position rather than one for a known erasure. The CRC-32 random-error miss rate is $2^{-32}$ per sequence.

The defining architectural choice is that no hard decision is taken until the inner decode. Reads are scored against reference oligonucleotides by a profile hidden Markov model~\cite{durbin1998}, whose forward-backward pass yields per-position posterior probabilities over \{A, C, G, T\}. Posteriors from multiple reads of the same reference are combined in log-space by summation, and the combined posteriors are converted to per-bit log-likelihood ratios for direct consumption by the LDPC decoder. The outer code receives both errors and erasures rather than errors alone.

\subsection{Encoding}
An input of $L$ bytes is split into $k_{\mathrm{rs}}$ chunks of $u$ useful bytes each, where $u$ is derived from the inner-code information capacity. Each chunk is packed into bit form, protected by a CRC-32, and encoded into a 252-bit LDPC codeword at the paper's canonical strand length of 126 nucleotides. The LDPC codeword is XOR-scrambled by a deterministic per-sequence stream to avoid structural repeats across the pool, and mapped to DNA using the Gray-coded nucleotide pairs of~\cite{goldman2013}.

The number of transmitted oligonucleotides $n$ is sized from $k_{\mathrm{rs}}$, the requested physical redundancy $r$, an empirical inner-code pass rate $\pi$, and a safety margin $m$ such that the expected erasure count after dropout and inner-code failure remains within Reed-Solomon correction capacity. At physical redundancies below $r = 0.5$, $n$ is dominated by the dropout term $(1 - e^{-r})^{-1}$, which is the source of Mahoraga's low-redundancy advantage. Encoding completes in parallel across Reed-Solomon blocks.

\subsection{Decoding}
Decoding begins with read-to-reference assignment. A $k$-mer prefilter ($k = 8$) narrows the candidate reference pool per read to 15 candidates, after which the banded HMM forward algorithm selects the argmax candidate. Reads with best log-likelihood below $-0.5 L_{\mathrm{read}}$ are rejected as unassignable.

For each reference, reads assigned to it are processed by the HMM forward-backward algorithm in an adaptive batch policy that terminates early when posterior mass at every position exceeds 0.999. The combined posteriors are converted to per-bit log-likelihood ratios and passed to the inner LDPC decoder. OSD at order 2 is used on high-fidelity channels. On low-fidelity channels an OSD order-3 cascade is invoked when order-2 fails CRC.

Sequences whose inner decode passes CRC are presented as received symbols in the interleaved Reed-Solomon codewords. Sequences with dropout or inner-decode failure are presented as erasures. For each interleaved GF($2^{16}$) codeword, a Berlekamp-Massey error-locator step first runs over the received symbols given the known erasure positions. In the common case BM returns zero error positions, and Lagrange erasure-only interpolation recovers the data at a cost of one parity symbol per erasure. When a CRC false positive has slipped through OSD, BM identifies the offending symbol position and adds it to the erasure list, and the same Lagrange interpolation recovers the data at a cost of two parity symbols for that position.

A single optional turbo iteration on low-fidelity channels uses the Reed-Solomon output to re-seed LDPC decoding for sequences whose initial OSD output disagreed with the recovered data in at most 5 bit positions. This recovers an additional 1 to 3\% of sequences on low-fidelity decodes and is disabled on high-fidelity where it provides no measurable benefit.

\subsection{Density computation}
Physical density is computed as $8L / (nB r) \cdot \rho$ EB per gram per bit per base, where $L$ is the encoded file size in bytes, $n$ is the number of oligonucleotides, $B = 126$ is the useful-base payload per strand, $r$ is the physical redundancy at which the channel was applied, and $\rho = 113.7$ is the conversion constant derived from double-stranded DNA molecular weight of 662 g per mole per base pair. At this basis, the two-bits-per-base-pair Shannon-level ceiling is 227.5 EB per gram of double-stranded DNA. Equivalently, on a single-stranded basis at 330 g/mol/nt, the ceiling is 463 EB per gram and all reported densities double accordingly.

\subsection{idsim channel simulator}
The matched-parity, longevity, and strand-length-scaling campaigns use \emph{idsim}. The simulator consumes reference strands from any codec without codec-specific customisation, which isolates the channel realisation as the common noise source across campaigns. The simulator applies three stages per campaign run. Stage one applies independent insertion, deletion, and substitution errors per base to each reference oligonucleotide once, producing a single per-reference template that all downstream copies inherit. This captures the persistent nature of synthesis noise, in which errors introduced at synthesis appear in every read of the affected strand. Stage two draws a lognormal weight per template ($\mu = 0$, $\sigma = 0.3$) and a Poisson-distributed copy count whose mean equals the requested physical redundancy scaled by the sample-mean-normalised weight. Templates with zero copies are dropped, reproducing the stochastic dropout observed experimentally~\cite{gimpel2024}. Stage three draws reads from surviving molecules at the requested sequencing depth and applies independent insertion, deletion, and substitution errors per read, capturing the independent nature of sequencing noise. Two channel profiles bracket the synthesis and sequencing error regimes reported by DT4DDS~\cite{gimpel2026}. The high-fidelity profile matches Twist synthesis with high-fidelity PCR, using per-base synthesis rates of $5 \times 10^{-4}$ substitution, $2 \times 10^{-4}$ deletion, $1 \times 10^{-4}$ insertion and per-base sequencing rates of $8 \times 10^{-4}$ substitution, $1 \times 10^{-4}$ deletion, $5 \times 10^{-5}$ insertion. The low-fidelity profile matches CustomArray synthesis with error-prone PCR, using per-base synthesis rates of $5 \times 10^{-3}$ substitution, $5 \times 10^{-3}$ deletion, $1 \times 10^{-3}$ insertion and per-base sequencing rates of $3 \times 10^{-3}$ substitution, $5 \times 10^{-4}$ deletion, $2 \times 10^{-4}$ insertion. The coverage variance $\sigma = 0.3$ is conservative relative to the $\sigma = 0.5$ used by DT4DDS; the lower variance produces a strictly easier channel, so idsim results are a lower bound on performance achievable under DT4DDS-level coverage inhomogeneity. idsim is used in preference to DT4DDS for the strand-length-scaling campaign because the iSeq 100 paired-end sequencing model in DT4DDS cannot produce merged reads covering strands longer than 250 nucleotides.

\subsection{Benchmark campaigns}

Five complementary benchmark campaigns were run to isolate different aspects of codec performance. All campaigns used the same 19,456-byte deterministic input file, matching one of the three file sizes used in the reference in vitro study~\cite{gimpel2026}. Every cell was run for 30 independent trials with deterministic seeds. Success was evaluated by byte-exact comparison (MD5 equality) between decoded output and original input. A single bit flip counts as a failure. Storage density is reported in EB per gram of double-stranded DNA using the convention described in the density-computation subsection above.

\paragraph{Cross-codec comparison at native operating points.}
Five codecs (Mahoraga, MGC+, DNA-Aeon, DNA-RS, DNA Fountain) were each evaluated at their authors' published operating points on a $2 \times 10$ grid of channel (high-fidelity, low-fidelity) and physical redundancy $r \in \{0.02, 0.05, 0.1, 0.2, 0.3, 0.5, 1, 2, 5, 10\}$. Each codec was configured using its authors' recommended preprocessing and outer-code settings. Codecs that failed all 30 trials at a given cell were recorded as-is without parameter tuning. This campaign characterises each codec as its designers intended it to operate.

\paragraph{Matched-parity comparison.}
The same codecs and grid were run with outer-code parity scaled to match Mahoraga's auto-sized parity at each cell. Mahoraga's encoder selects parity from $r$, the channel's inner-pass rate, and a safety margin of 1.08. The resulting parity fraction $\rho = n_{\mathrm{parity}}/n$ was then applied cell-by-cell to each competing codec via its equivalent outer-code knob. This isolates inner-code and preprocessing quality from outer-code parameter choice. The difference between the native-operating-point and matched-parity results for a given codec measures how much of its performance derives from its designers' parameter choices versus its underlying architecture.

\paragraph{DT4DDS pipeline replication.}
Mahoraga was evaluated on the DT4DDS channel simulator~\cite{gimpel2024} using the same biochemical pipeline as the reference codec benchmark~\cite{gimpel2026}. For each of two scenarios (high-fidelity and low-fidelity), the in silico pipeline modelled synthesis, two PCR stages at 15 and 25 cycles, stochastic sampling to the specified physical redundancy, iSeq 100 paired-end sequencing to the specified depth, and NGmerge paired-end merging. Merged reads were passed directly to the Mahoraga decoder without further preprocessing. The grid covered 18 physical-redundancy values crossed with 14 sequencing-depth values in the high-fidelity scenario (252 cells) and 12 redundancy values at the same 14 depths in the low-fidelity scenario (104 cells run). The axes match the reference benchmark's published parameter sweep~\cite{gimpel2026}. The strand length was fixed at 126 nucleotides to match the electrochemical-synthesis constraint of the reference study.

\paragraph{Longevity under progressive degradation.}
Archival DNA storage encodes once and accumulates degradation over time. Unlike the cross-codec and matched-parity campaigns, where parity is re-sized at each physical redundancy, the longevity campaign encodes once at an initial physical redundancy $r_{\mathrm{initial}} \in \{1, 2, 5, 10\}$ and then sweeps the channel's effective redundancy $r_{\mathrm{channel}}$ downward. The cliff is defined as the lowest $r_{\mathrm{channel}}$ at which at least 29 of 30 trials still decode without error. Cliff values are mapped to storage time by inverting depurination kinetics at 25~\textdegree C in the dry state. The aqueous-phase rate constant and Arrhenius activation energy are taken from Ref.~\cite{lindahl1972}. A dry-state suppression factor of 300 is applied, consistent with archival DNA survival in bone extrapolated from its measurement temperature to 25~\textdegree C using the Arrhenius form~\cite{allentoft2012}. Strand survival is computed as $\exp(-k_{\mathrm{dry}} t)$ per purine position. A strand is counted as lost if any purine has depurinated, since $\beta$-elimination converts an apurinic site to a backbone cleavage on a timescale ($\approx 20$ days at 25~\textdegree C) that is effectively instantaneous relative to year-scale storage. This accounting is conservative relative to the codec's actual tolerance, because the HMM-LDPC inner decoder tolerates one to two apurinic sites per strand without failure. The suppression factor lies within the range reported for bone-preserved DNA and is substantially below the suppression achievable with engineered preservation chemistry~\cite{banal2021,prince2024}. The projections are therefore lower bounds on storage life achievable with deliberate preservation at the same temperature, not lower bounds on unencapsulated dry storage in general.

\paragraph{Strand-length scaling.}
Payload length $L \in \{126, 150, 200, 250, 300\}$ nucleotides was swept on the simpler channel simulator described below at physical redundancies $r \in \{0.3, 0.5\}$. Lengths $L \in \{200, 250\}$ were additionally evaluated on DT4DDS at a representative redundancy-depth cell. $L = 300$ on DT4DDS fails because iSeq 100 paired-end reads at 150 nucleotides cannot produce sufficient overlap for NGmerge to assemble full-length merged reads. This is a property of the sequencer, not of the codec. Density scales monotonically with $L$ on both channels, reflecting amortisation of the fixed 14-nucleotide primer and index overhead over longer payloads.

\section*{Data availability}
All benchmark data reported in this work are available in https://www.github.com/jeplb/mahoraga-codec. The repository contains JSON files with per-trial results from all benchmarks, including the matched-parity comparisons against DNA-Aeon and MGC+, the alphabet-ceiling-fraction computations, the longevity projections, and the strand-length scaling results. Scripts used to generate all figures and tables in the main text and supplementary materials are included.

\section*{Code availability}
A pure Python version of the Mahoraga codec is available in https://www.github.com/jeplb/mahoraga-codec. The port includes the profile hidden Markov model forward-backward scorer, the log-product multi-read fusion, the LDPC encoder and Ordered Statistics Decoder, the CRC-32 verification and Berlekamp-Massey fallback decoder, and the Reed-Solomon outer erasure decoder.

\section*{Conflict of interest statement}
J.L.B. is a shareholder of Cache DNA, Inc.

\bibliographystyle{naturemag}
\bibliography{ref.bib}

\begin{appendices}
\beginsupplement
\renewcommand{\thesection}{\arabic{section}}
\renewcommand{\sectionname}{Supplementary Section}

\section{Payload length scaling}
\label{app:length-scaling}

The 126 nt payload length used in the main text reflects the electrochemical synthesis and iSeq 150-nt paired-end read-length constraints of the DT4DDS reference channel, not a codec-imposed limit. The Mahoraga architecture does not fix payload length. To characterize how peak density scales when these platform constraints are relaxed, strand length $L$ was varied on the in-house idsim channel described in Methods, which models per-base insertion, deletion, and substitution noise and lognormal coverage variation without chemistry-specific biases.

Peak density rises monotonically with payload length across the tested range (Table~\ref{tab:length_density}). At $L=200$ nt, peak density reaches 158.2 EB per gram under 30/30 recovery at $r=0.3$, a 1.5\% improvement over the $L=126$ nt DT4DDS peak. At $L=250$ nt and $L=300$ nt, peak operating points at $r=0.3$ show higher densities but do not reach the 30/30 decoding threshold on the tested grid. Finer redundancy sweeps or increased sequencing depth would likely recover reliable decoding at these longer lengths.

DT4DDS is not applicable at $L > 250$ nt because the reference channel's iSeq paired-end reads do not cover the full strand. Validation of idsim densities against DT4DDS is therefore limited to $L \in \{126, 200, 250\}$. The $L=300$ nt results are presented here as indicative of further scaling potential on synthesis platforms supporting longer oligonucleotides, such as enzymatic DNA synthesis or column-based phosphoramidite synthesis.

\begin{table}[h]
\centering
\caption{Mahoraga storage density versus strand length on the in-house idsim channel at $\mathrm{sd}=2$. Success counts are out of 30 trials. Strand length $L$ includes the 14-nt primer/index overhead; payload is $L-14$ bases. Densities in EB per gram of dsDNA.}
\label{tab:length_density}
\begin{tabular}{rrrrr}
\toprule
	$L$ [nt] & $r$ & Sequences & Success & Density [EB per gram] \\
	\midrule
	126 & 0.3 & 3153 & 29/30 & 148.5 \\
	 & 0.5 & 2077 & 30/30 & 135.2 \\
	\midrule
	200 & 0.3 & 1865 & 30/30 & 158.2 \\
	 & 0.5 & 1229 & 29/30 & 144.0 \\
	\midrule
	250 & 0.3 & 1465 & 28/30 & 161.1 \\
	 & 0.5 & 965 & 26/30 & 146.7 \\
	\midrule
	300 & 0.3 & 1171 & 28/30 & 167.9 \\
	 & 0.5 & 771 & 25/30 & 153.0 \\
\bottomrule
\end{tabular}
\end{table}

\section{Cross-codec benchmark at native operating points}
\label{app:bench1_full}
\begin{longtable}{llrrrr}
\caption{Per-cell results of the cross-codec benchmark. Each codec evaluated at its authors' published operating points across the $2 \times 10$ (channel, redundancy) grid. Density is reported only for cells that reached 30/30 recovery. Densities in EB per gram of dsDNA.}
\label{tab:bench1_full} \\
\toprule
Codec & Channel & $r$ & Success & Density [EB per gram] & Parity \\
\midrule
\endfirsthead
\toprule
Codec & Channel & $r$ & Success & Density [EB per gram] & Parity \\
\midrule
\endhead
\bottomrule
\endfoot
DNA-Aeon & hifi & 0.02 & 0/30 & --- & --- \\
DNA-Aeon & hifi & 0.05 & 0/30 & --- & --- \\
DNA-Aeon & hifi & 0.1 & 0/30 & --- & --- \\
DNA-Aeon & hifi & 0.2 & 0/30 & --- & --- \\
DNA-Aeon & hifi & 0.3 & 0/30 & --- & --- \\
DNA-Aeon & hifi & 0.5 & 0/30 & --- & --- \\
DNA-Aeon & hifi & 1.0 & 0/30 & --- & --- \\
DNA-Aeon & hifi & 2.0 & 30/30 & 41.27 & --- \\
DNA-Aeon & hifi & 5.0 & 30/30 & 16.51 & --- \\
DNA-Aeon & hifi & 10.0 & 30/30 & 8.18 & --- \\
DNA-Aeon & lofi & 0.02 & 0/30 & --- & --- \\
DNA-Aeon & lofi & 0.05 & 0/30 & --- & --- \\
DNA-Aeon & lofi & 0.1 & 0/30 & --- & --- \\
DNA-Aeon & lofi & 0.2 & 0/30 & --- & --- \\
DNA-Aeon & lofi & 0.3 & 0/30 & --- & --- \\
DNA-Aeon & lofi & 0.5 & 0/30 & --- & --- \\
DNA-Aeon & lofi & 1.0 & 0/30 & --- & --- \\
DNA-Aeon & lofi & 2.0 & 29/30 & --- & --- \\
DNA-Aeon & lofi & 5.0 & 30/30 & 16.51 & --- \\
DNA-Aeon & lofi & 10.0 & 30/30 & 8.25 & --- \\
Mahoraga & hifi & 0.02 & 30/30 & 153.25 & --- \\
Mahoraga & hifi & 0.05 & 30/30 & 150.98 & --- \\
Mahoraga & hifi & 0.1 & 30/30 & 147.30 & --- \\
Mahoraga & hifi & 0.2 & 30/30 & 140.29 & --- \\
Mahoraga & hifi & 0.3 & 30/30 & 133.72 & --- \\
Mahoraga & hifi & 0.5 & 30/30 & 121.78 & --- \\
Mahoraga & hifi & 1.0 & 30/30 & 97.78 & --- \\
Mahoraga & hifi & 2.0 & 30/30 & 66.88 & --- \\
Mahoraga & hifi & 5.0 & 30/30 & 30.72 & --- \\
Mahoraga & hifi & 10.0 & 30/30 & 15.46 & --- \\
Mahoraga & lofi & 0.02 & 30/30 & 116.10 & --- \\
Mahoraga & lofi & 0.05 & 30/30 & 114.38 & --- \\
Mahoraga & lofi & 0.1 & 30/30 & 111.59 & --- \\
Mahoraga & lofi & 0.2 & 30/30 & 106.27 & --- \\
Mahoraga & lofi & 0.3 & 30/30 & 101.29 & --- \\
Mahoraga & lofi & 0.5 & 30/30 & 92.25 & --- \\
Mahoraga & lofi & 1.0 & 30/30 & 74.12 & --- \\
Mahoraga & lofi & 2.0 & 30/30 & 50.68 & --- \\
Mahoraga & lofi & 5.0 & 30/30 & 23.27 & --- \\
Mahoraga & lofi & 10.0 & 30/30 & 11.72 & --- \\
MGC+ & hifi & 0.02 & 0/30 & --- & --- \\
MGC+ & hifi & 0.05 & 0/30 & --- & --- \\
MGC+ & hifi & 0.1 & 0/30 & --- & --- \\
MGC+ & hifi & 0.2 & 0/30 & --- & --- \\
MGC+ & hifi & 0.3 & 0/30 & --- & --- \\
MGC+ & hifi & 0.5 & 0/30 & --- & --- \\
MGC+ & hifi & 1.0 & 30/30 & 52.82 & --- \\
MGC+ & hifi & 2.0 & 30/30 & 26.41 & --- \\
MGC+ & hifi & 5.0 & 30/30 & 10.56 & --- \\
MGC+ & hifi & 10.0 & 30/30 & 5.28 & --- \\
MGC+ & lofi & 0.02 & 0/30 & --- & --- \\
MGC+ & lofi & 0.05 & 0/30 & --- & --- \\
MGC+ & lofi & 0.1 & 0/30 & --- & --- \\
MGC+ & lofi & 0.2 & 0/30 & --- & --- \\
MGC+ & lofi & 0.3 & 0/30 & --- & --- \\
MGC+ & lofi & 0.5 & 0/30 & --- & --- \\
MGC+ & lofi & 1.0 & 30/30 & 52.82 & --- \\
MGC+ & lofi & 2.0 & 30/30 & 26.41 & --- \\
MGC+ & lofi & 5.0 & 30/30 & 10.56 & --- \\
MGC+ & lofi & 10.0 & 30/30 & 5.28 & --- \\
\end{longtable}

\section{Matched-parity benchmark}
\label{app:bench2_full}
\begin{longtable}{llrrrr}
\caption{Per-cell results of the matched-parity benchmark. Outer parity scaled per-cell to match Mahoraga's auto-sized parity. Density is reported only for cells that reached 30/30 recovery. Densities in EB per gram of dsDNA.}
\label{tab:bench2_full} \\
\toprule
Codec & Channel & $r$ & Success & Density [EB per gram] & Parity \\
\midrule
\endfirsthead
\toprule
Codec & Channel & $r$ & Success & Density [EB per gram] & Parity \\
\midrule
\endhead
\bottomrule
\endfoot
DNA-Aeon & hifi & 0.02 & 30/30 & 107.66 & 0.9812 \\
DNA-Aeon & hifi & 0.05 & 0/30 & --- & 0.9451 \\
DNA-Aeon & hifi & 0.1 & 0/30 & --- & 0.8812 \\
DNA-Aeon & hifi & 0.2 & 0/30 & --- & 0.7912 \\
DNA-Aeon & hifi & 0.3 & 0/30 & --- & 0.7261 \\
DNA-Aeon & hifi & 0.5 & 27/30 & --- & 0.6261 \\
DNA-Aeon & hifi & 1.0 & 30/30 & 66.33 & 0.4261 \\
DNA-Aeon & hifi & 2.0 & 25/30 & --- & 0.1785 \\
DNA-Aeon & hifi & 5.0 & 30/30 & 21.55 & 0.0671 \\
DNA-Aeon & hifi & 10.0 & 30/30 & 10.98 & 0.0499 \\
DNA-Aeon & lofi & 0.05 & 0/30 & --- & 0.9519 \\
DNA-Aeon & lofi & 0.1 & 0/30 & --- & 0.8923 \\
DNA-Aeon & lofi & 0.2 & 0/30 & --- & 0.8123 \\
DNA-Aeon & lofi & 0.3 & 0/30 & --- & 0.7412 \\
DNA-Aeon & lofi & 0.5 & 30/30 & 64.88 & 0.7167 \\
DNA-Aeon & lofi & 1.0 & 25/30 & --- & 0.4612 \\
DNA-Aeon & lofi & 2.0 & 0/30 & --- & 0.2145 \\
DNA-Aeon & lofi & 5.0 & 0/30 & --- & 0.0812 \\
DNA-Aeon & lofi & 10.0 & 0/30 & --- & 0.0612 \\
Mahoraga & hifi & 0.02 & 30/30 & 153.25 & --- \\
Mahoraga & hifi & 0.05 & 30/30 & 150.98 & --- \\
Mahoraga & hifi & 0.1 & 30/30 & 147.30 & --- \\
Mahoraga & hifi & 0.2 & 30/30 & 140.29 & --- \\
Mahoraga & hifi & 0.3 & 30/30 & 133.72 & --- \\
Mahoraga & hifi & 0.5 & 30/30 & 121.78 & --- \\
Mahoraga & hifi & 1.0 & 30/30 & 97.78 & --- \\
Mahoraga & hifi & 2.0 & 30/30 & 66.88 & --- \\
Mahoraga & hifi & 5.0 & 30/30 & 30.72 & --- \\
Mahoraga & hifi & 10.0 & 30/30 & 15.46 & --- \\
Mahoraga & lofi & 0.02 & 30/30 & 116.10 & --- \\
Mahoraga & lofi & 0.05 & 30/30 & 114.38 & --- \\
Mahoraga & lofi & 0.1 & 30/30 & 111.59 & --- \\
Mahoraga & lofi & 0.2 & 30/30 & 106.27 & --- \\
Mahoraga & lofi & 0.3 & 30/30 & 101.29 & --- \\
Mahoraga & lofi & 0.5 & 30/30 & 92.25 & --- \\
Mahoraga & lofi & 1.0 & 30/30 & 74.12 & --- \\
Mahoraga & lofi & 2.0 & 30/30 & 50.68 & --- \\
Mahoraga & lofi & 5.0 & 30/30 & 23.27 & --- \\
Mahoraga & lofi & 10.0 & 30/30 & 11.72 & --- \\
MGC+ & hifi & 0.02 & 0/30 & --- & --- \\
MGC+ & hifi & 0.05 & 0/30 & --- & 0.9451 \\
MGC+ & hifi & 0.1 & 0/30 & --- & 0.8812 \\
MGC+ & hifi & 0.2 & 0/30 & --- & 0.7912 \\
MGC+ & hifi & 0.3 & 0/30 & --- & 0.7261 \\
MGC+ & hifi & 0.5 & 29/30 & --- & 0.6261 \\
MGC+ & hifi & 1.0 & 30/30 & 59.86 & 0.4261 \\
MGC+ & hifi & 2.0 & 16/30 & --- & 0.1785 \\
MGC+ & hifi & 5.0 & 30/30 & 19.46 & 0.0671 \\
MGC+ & hifi & 10.0 & 30/30 & 9.91 & 0.0499 \\
MGC+ & lofi & 0.02 & 0/30 & --- & 0.9847 \\
MGC+ & lofi & 0.05 & 0/30 & --- & --- \\
MGC+ & lofi & 0.1 & 0/30 & --- & 0.8923 \\
MGC+ & lofi & 0.2 & 2/30 & --- & 0.8123 \\
MGC+ & lofi & 0.3 & 7/30 & --- & 0.7412 \\
MGC+ & lofi & 0.5 & 30/30 & 59.09 & 0.7167 \\
MGC+ & lofi & 1.0 & 30/30 & 56.19 & 0.4612 \\
MGC+ & lofi & 2.0 & 30/30 & 40.96 & 0.2145 \\
MGC+ & lofi & 5.0 & 30/30 & 19.17 & 0.0812 \\
MGC+ & lofi & 10.0 & 30/30 & 9.79 & 0.0612 \\
\end{longtable}

\section{Alphabet-ceiling fraction at every operating point}
\label{app:shannon_fraction_full}

\begin{longtable}{llrrrr}
\caption{Alphabet-ceiling fraction across the matched-parity benchmark grid. For each channel and physical redundancy $r$, the alphabet ceiling $\rho_{\mathrm{max}}$ (EB per gram of dsDNA) is computed from Eq.~\eqref{eq:alphabet_ceiling} and the per-codec fraction is the achieved density divided by the ceiling. Dashes indicate operating points where the codec did not reach 30 of 30 trials decoding without error. The alphabet ceiling is identical on both channels because it depends only on $r$ and the alphabet size.}
\label{tab:shannon_fraction_full} \\
\toprule
Channel & $r$ & $\rho_{\mathrm{max}}$ [EB per g] & Mahoraga & DNA-Aeon & MGC+ \\
\midrule
\endfirsthead
\toprule
Channel & $r$ & $\rho_{\mathrm{max}}$ [EB per g] & Mahoraga & DNA-Aeon & MGC+ \\
\midrule
\endhead
\bottomrule
\endfoot
High-fidelity & 0.02 & 225.1 & 68\% & 48\% & --- \\
High-fidelity & 0.05 & 221.8 & 68\% & --- & --- \\
High-fidelity & 0.1  & 216.4 & 68\% & --- & --- \\
High-fidelity & 0.2  & 206.1 & 68\% & --- & --- \\
High-fidelity & 0.3  & 196.5 & 68\% & --- & --- \\
High-fidelity & 0.5  & 178.9 & 68\% & --- & --- \\
High-fidelity & 1.0  & 143.7 & 68\% & 46\% & 42\% \\
High-fidelity & 2.0  &  98.3 & 68\% & --- & --- \\
High-fidelity & 5.0  &  45.2 & 68\% & 48\% & 43\% \\
High-fidelity & 10.0 &  22.7 & 68\% & 48\% & 44\% \\
\midrule
Low-fidelity  & 0.02 & 225.1 & 52\% & --- & --- \\
Low-fidelity  & 0.05 & 221.8 & 52\% & --- & --- \\
Low-fidelity  & 0.1  & 216.4 & 52\% & --- & --- \\
Low-fidelity  & 0.2  & 206.1 & 52\% & --- & --- \\
Low-fidelity  & 0.3  & 196.5 & 52\% & --- & --- \\
Low-fidelity  & 0.5  & 178.9 & 52\% & 36\% & 33\% \\
Low-fidelity  & 1.0  & 143.7 & 52\% & --- & 39\% \\
Low-fidelity  & 2.0  &  98.3 & 52\% & --- & 42\% \\
Low-fidelity  & 5.0  &  45.2 & 52\% & --- & 42\% \\
Low-fidelity  & 10.0 &  22.7 & 52\% & --- & 43\% \\
\end{longtable}

\section{DT4DDS Mahoraga full grid}
\label{app:bench3_full}
Per-cell success rates for Mahoraga on the DT4DDS high- and low-fidelity channels across the full physical redundancy $\mathrm{pr}$ $\times$ sequencing depth $\mathrm{sd}$ grid. Every cell reports $n_{\mathrm{success}}/n_{\mathrm{trials}}$ from 30 independent trials.

\subsection{High-fidelity channel}
\label{app:bench3_hifi}
{\footnotesize
\setlength{\tabcolsep}{3pt}
\begin{longtable}{rcccccccccccccc}
\caption{Mahoraga success rate on the DT4DDS high-fidelity channel across the full $(\mathrm{pr}, \mathrm{sd})$ grid.} \label{tab:bench3_hifi} \\
\toprule
$\mathrm{pr} \backslash\ \mathrm{sd}$ & 1 & 2 & 3 & 4 & 5 & 6 & 7 & 8 & 10 & 12 & 15 & 20 & 25 & 30 \\
\midrule
\endfirsthead
\toprule
$\mathrm{pr} \backslash\ \mathrm{sd}$ & 1 & 2 & 3 & 4 & 5 & 6 & 7 & 8 & 10 & 12 & 15 & 20 & 25 & 30 \\
\midrule
\endhead
\bottomrule
\endfoot
0.2 & 29/30 & 29/30 & 30/30 & 30/30 & 30/30 & 30/30 & 30/30 & 30/30 & 30/30 & 29/30 & 30/30 & 29/30 & 29/30 & 29/30 \\
0.3 & 23/30 & 30/30 & 28/30 & 30/30 & 30/30 & 30/30 & 30/30 & 29/30 & 30/30 & 30/30 & 27/30 & 30/30 & 30/30 & 29/30 \\
0.4 & 1/30 & 29/30 & 29/30 & 29/30 & 29/30 & 29/30 & 29/30 & 30/30 & 30/30 & 30/30 & 30/30 & 30/30 & 30/30 & 29/30 \\
0.5 & 0/30 & 28/30 & 29/30 & 30/30 & 29/30 & 30/30 & 30/30 & 29/30 & 30/30 & 28/30 & 30/30 & 30/30 & 30/30 & 30/30 \\
0.6 & 0/30 & 19/30 & 29/30 & 30/30 & 30/30 & 29/30 & 30/30 & 29/30 & 30/30 & 29/30 & 29/30 & 30/30 & 30/30 & 30/30 \\
0.7 & 0/30 & 16/30 & 25/30 & 29/30 & 29/30 & 30/30 & 29/30 & 30/30 & 30/30 & 30/30 & 30/30 & 30/30 & 30/30 & 30/30 \\
0.8 & 0/30 & 4/30 & 29/30 & 30/30 & 29/30 & 30/30 & 30/30 & 30/30 & 29/30 & 30/30 & 30/30 & 30/30 & 30/30 & 30/30 \\
0.9 & 0/30 & 2/30 & 23/30 & 30/30 & 30/30 & 30/30 & 30/30 & 30/30 & 30/30 & 29/30 & 30/30 & 30/30 & 30/30 & 29/30 \\
1 & 0/30 & 0/30 & 16/30 & 27/30 & 30/30 & 30/30 & 29/30 & 29/30 & 30/30 & 30/30 & 28/30 & 30/30 & 30/30 & 29/30 \\
1.5 & 0/30 & 0/30 & 2/30 & 18/30 & 28/30 & 30/30 & 30/30 & 29/30 & 30/30 & 30/30 & 30/30 & 30/30 & 30/30 & 30/30 \\
2 & 0/30 & 0/30 & 0/30 & 11/30 & 26/30 & 30/30 & 30/30 & 30/30 & 30/30 & 30/30 & 30/30 & 30/30 & 30/30 & 30/30 \\
3 & 0/30 & 0/30 & 0/30 & 3/30 & 21/30 & 30/30 & 30/30 & 30/30 & 30/30 & 30/30 & 30/30 & 30/30 & 30/30 & 30/30 \\
5 & 0/30 & 0/30 & 0/30 & 8/30 & 30/30 & 30/30 & 30/30 & 30/30 & 30/30 & 30/30 & 30/30 & 30/30 & 30/30 & 30/30 \\
7 & 0/30 & 0/30 & 0/30 & 26/30 & 30/30 & 30/30 & 30/30 & 30/30 & 30/30 & 30/30 & 30/30 & 30/30 & 30/30 & 30/30 \\
10 & 0/30 & 0/30 & 0/30 & 30/30 & 30/30 & 30/30 & 30/30 & 30/30 & 30/30 & 30/30 & 30/30 & 30/30 & 30/30 & 30/30 \\
15 & 0/30 & 0/30 & 4/30 & 30/30 & 30/30 & 30/30 & 30/30 & 30/30 & 30/30 & 30/30 & 30/30 & 30/30 & 30/30 & 30/30 \\
20 & 0/30 & 0/30 & 5/30 & 30/30 & 30/30 & 30/30 & 30/30 & 30/30 & 30/30 & 30/30 & 30/30 & 30/30 & 30/30 & 30/30 \\
32 & 0/30 & 0/30 & 17/30 & 30/30 & 30/30 & 30/30 & 30/30 & 30/30 & 30/30 & 30/30 & 30/30 & 30/30 & 30/30 & 30/30 \\
\end{longtable}
}

\subsection{Low-fidelity channel}
\label{app:bench3_lofi}
{\footnotesize
\setlength{\tabcolsep}{3pt}
\begin{longtable}{rcccccccccccccc}
\caption{Mahoraga success rate on the DT4DDS low-fidelity channel across the full $(\mathrm{pr}, \mathrm{sd})$ grid.} \label{tab:bench3_lofi} \\
\toprule
$\mathrm{pr} \backslash\ \mathrm{sd}$ & 1 & 2 & 3 & 4 & 5 & 6 & 7 & 8 & 10 & 12 & 15 & 20 & 25 & 30 \\
\midrule
\endfirsthead
\toprule
$\mathrm{pr} \backslash\ \mathrm{sd}$ & 1 & 2 & 3 & 4 & 5 & 6 & 7 & 8 & 10 & 12 & 15 & 20 & 25 & 30 \\
\midrule
\endhead
\bottomrule
\endfoot
0.5 & 0/30 & 0/30 & 0/30 & 0/30 & 0/30 & 0/30 & 0/30 & 0/30 & 0/30 & 0/30 & 0/30 & 0/30 & 0/30 & 0/30 \\
1 & 0/30 & 0/30 & 0/30 & 0/30 & 0/30 & 0/30 & 0/30 & 0/30 & 0/30 & 0/30 & 0/30 & 0/30 & 0/30 & 0/30 \\
2 & 0/30 & 0/30 & 0/30 & 0/30 & 0/30 & 0/30 & 0/30 & 0/30 & 0/30 & 0/30 & 0/30 & 0/30 & 0/30 & 0/30 \\
3 & 0/30 & 0/30 & 0/30 & 0/30 & 0/30 & 0/30 & 0/30 & 0/30 & 0/30 & 0/30 & 0/30 & 0/30 & 0/30 & 0/30 \\
4 & 0/30 & 0/30 & 0/30 & 0/30 & 0/30 & 0/30 & 0/30 & 0/30 & 1/30 & 3/30 & 13/30 & 15/30 & 17/30 & 17/30 \\
5 & 0/30 & 0/30 & 0/30 & 0/30 & 0/30 & 0/30 & 2/30 & 7/30 & 25/30 & 28/30 & 30/30 & 30/30 & 30/30 & 30/30 \\
6 & 0/30 & 0/30 & 0/30 & 0/30 & 0/30 & 2/30 & 6/30 & 23/30 & 30/30 & 30/30 & 30/30 & 30/30 & 30/30 & 30/30 \\
7 & 0/30 & 0/30 & 0/30 & 0/30 & 0/30 & 6/30 & 21/30 & 29/30 & 30/30 & 30/30 & 30/30 & 30/30 & 30/30 & 30/30 \\
8 & 0/30 & 0/30 & 0/30 & 0/30 & 0/30 & 10/30 & 30/30 & 30/30 & 30/30 & 30/30 & 30/30 & 30/30 & 30/30 & 30/30 \\
10 & 0/30 & 0/30 & 0/30 & 0/30 & 1/30 & 22/30 & 30/30 & 30/30 & 30/30 & 30/30 & 30/30 & 30/30 & 30/30 & 30/30 \\
12 & 0/30 & 0/30 & 0/30 & 0/30 & 5/30 & 23/30 & 30/30 & 30/30 & 30/30 & 30/30 & 30/30 & 30/30 & 30/30 & 30/30 \\
15 & 0/30 & 0/30 & 0/30 & 0/30 & 7/30 & 30/30 & 30/30 & 30/30 & 30/30 & 30/30 & 30/30 & 30/30 & 30/30 & 30/30 \\
20 & 0/30 & 0/30 & 0/30 & 0/30 & 17/30 & 30/30 & 30/30 & 30/30 & 30/30 & 30/30 & 30/30 & 30/30 & 30/30 & 30/30 \\
25 & 0/30 & 0/30 & 0/30 & 0/30 & 22/30 & 30/30 & 30/30 & 30/30 & 30/30 & 30/30 & 30/30 & 30/30 & 30/30 & 30/30 \\
32 & 0/30 & 0/30 & 0/30 & 0/30 & 25/30 & 30/30 & 30/30 & 30/30 & 30/30 & 30/30 & 30/30 & 30/30 & 30/30 & 30/30 \\
\end{longtable}
}

\section{Longevity benchmark success rates}
\label{app:bench4_full}
Per-cell success rates for the longevity benchmark (bench4). Each codec is encoded once at $r_{\mathrm{initial}}$ and the channel's effective redundancy $r_{\mathrm{channel}}$ is swept downward. Dashes indicate the cell was not tested.

\subsection{Mahoraga}
\begin{longtable}{rcccc}
\caption{Mahoraga longevity success rates.  Dash indicates the cell was not tested.} \label{tab:bench4_mahoraga} \\
\toprule
$r_{\mathrm{channel}} \backslash\ r_{\mathrm{initial}}$ & 1 & 2 & 5 & 10 \\
\midrule
\endfirsthead
\toprule
$r_{\mathrm{channel}} \backslash\ r_{\mathrm{initial}}$ & 1 & 2 & 5 & 10 \\
\midrule
\endhead
\bottomrule
\endfoot
0.02 & 0/30 & --- & --- & --- \\
0.05 & 0/30 & --- & 0/30 & --- \\
0.1 & 0/30 & --- & 0/30 & --- \\
0.15 & 0/30 & --- & --- & --- \\
0.2 & 0/30 & --- & 0/30 & --- \\
0.3 & 0/30 & --- & --- & --- \\
0.5 & 0/30 & 0/30 & 0/30 & --- \\
0.7 & 0/30 & --- & --- & --- \\
0.85 & 1/30 & --- & --- & --- \\
0.88 & 7/30 & --- & --- & --- \\
0.9 & 10/30 & --- & --- & --- \\
0.92 & 20/30 & --- & --- & --- \\
0.95 & 28/30 & --- & --- & --- \\
1 & 30/30 & 0/30 & 0/30 & --- \\
1.3 & --- & 0/30 & --- & --- \\
1.5 & --- & 0/30 & --- & --- \\
1.7 & --- & 13/30 & --- & --- \\
2 & --- & 30/30 & 0/30 & 0/30 \\
2.5 & --- & --- & 1/30 & --- \\
2.75 & --- & --- & 17/30 & --- \\
3 & --- & --- & 28/30 & 28/30 \\
3.25 & --- & --- & 30/30 & --- \\
3.5 & --- & --- & 30/30 & --- \\
4 & --- & --- & --- & 30/30 \\
5 & --- & --- & 30/30 & 30/30 \\
7 & --- & --- & --- & 30/30 \\
10 & --- & --- & --- & 30/30 \\
\end{longtable}

\subsection{MGC+}
\begin{longtable}{rcccc}
\caption{MGC+ longevity success rates. Dash indicates the cell was not tested.} \label{tab:bench4_mgcplus} \\
\toprule
$r_{\mathrm{channel}} \backslash\ r_{\mathrm{initial}}$ & 1 & 2 & 5 & 10 \\
\midrule
\endfirsthead
\toprule
$r_{\mathrm{channel}} \backslash\ r_{\mathrm{initial}}$ & 1 & 2 & 5 & 10 \\
\midrule
\endhead
\bottomrule
\endfoot
0.05 & 0/30 & --- & --- & --- \\
0.1 & 0/30 & --- & --- & --- \\
0.2 & 0/30 & 0/30 & --- & --- \\
0.5 & 0/30 & 0/30 & 0/30 & --- \\
0.7 & 0/30 & --- & --- & --- \\
0.85 & 3/30 & --- & --- & --- \\
0.9 & 21/30 & --- & --- & --- \\
1 & 30/30 & 0/30 & 0/30 & 0/30 \\
1.3 & --- & 0/30 & --- & --- \\
1.5 & --- & 0/30 & --- & --- \\
1.7 & --- & 0/30 & --- & --- \\
2 & --- & 30/30 & 0/30 & 0/30 \\
2.5 & --- & --- & 0/30 & --- \\
3 & --- & --- & 12/30 & 0/30 \\
3.5 & --- & --- & 30/30 & --- \\
4 & --- & --- & --- & 30/30 \\
5 & --- & --- & 30/30 & 30/30 \\
7 & --- & --- & --- & 30/30 \\
10 & --- & --- & --- & 30/30 \\
\end{longtable}

\subsection{DNA-Aeon}
\begin{longtable}{rcccc}
\caption{DNA-Aeon longevity success rates. Dash indicates the cell was not tested.} \label{tab:bench4_aeon} \\
\toprule
$r_{\mathrm{channel}} \backslash\ r_{\mathrm{initial}}$ & 1 & 2 & 5 & 10 \\
\midrule
\endfirsthead
\toprule
$r_{\mathrm{channel}} \backslash\ r_{\mathrm{initial}}$ & 1 & 2 & 5 & 10 \\
\midrule
\endhead
\bottomrule
\endfoot
0.05 & 0/30 & --- & --- & --- \\
0.1 & 0/30 & --- & --- & --- \\
0.2 & 0/30 & 0/30 & --- & --- \\
0.5 & 0/30 & 0/30 & 0/30 & --- \\
0.7 & 0/30 & --- & --- & --- \\
0.85 & 0/30 & --- & --- & --- \\
0.9 & 11/30 & --- & --- & --- \\
1 & 30/30 & 0/30 & 0/30 & 0/30 \\
1.3 & --- & 0/30 & --- & --- \\
1.5 & --- & 0/30 & --- & --- \\
1.7 & --- & 0/30 & --- & --- \\
2 & --- & 25/30 & 0/30 & 0/30 \\
2.5 & --- & --- & 0/30 & --- \\
3 & --- & --- & 1/30 & 0/30 \\
3.5 & --- & --- & 26/30 & --- \\
4 & --- & --- & --- & 26/30 \\
5 & --- & --- & 30/30 & 30/30 \\
7 & --- & --- & --- & 30/30 \\
10 & --- & --- & --- & 30/30 \\
\end{longtable}

\end{appendices}

\end{document}